\documentclass[showpacs,aps,prd,reprint,superscriptaddress,nofootinbib,longbibliography]{revtex4-1}
\usepackage[colorlinks=true, pdfstartview=FitV, bookmarks=true, bookmarksnumbered=true, breaklinks]{hyperref}
\usepackage[dvipdfmx]{graphicx}
\usepackage{amsmath,amssymb,bm,color,longtable,mathrsfs,amsfonts,slashed}

\definecolor{DeepPink2}{rgb}{0.932,0.07,0.536}
\definecolor{RoyalBlue1}{rgb}{0.284,0.464,1}
\definecolor{SpringGreen3}{rgb}{0,0.804,0.4}  
\hypersetup{linkcolor=RoyalBlue1, citecolor=DeepPink2, urlcolor=SpringGreen3}

\begin{document}
\begin{flushright}
\end{flushright}

\title{Spectrum of singly heavy baryons from a chiral effective theory of diquarks}

\author{Yonghee~Kim}
\email[]{\tt kimu.ryonhi@phys.kyushu-u.ac.jp}
\affiliation{Department of Physics, Kyushu University, Fukuoka 819-0395, Japan}

\author{Emiko~Hiyama}
\email[]{\tt hiyama@riken.jp}
\affiliation{Department of Physics, Kyushu University, Fukuoka 819-0395, Japan}
\affiliation{Nishina Center for Accelerator-Based Science, RIKEN, Wako 351-0198, Japan}

\author{Makoto~Oka}
\email[]{\tt oka@post.j-parc.jp}
\affiliation{Advanced Science Research Center, Japan Atomic Energy Agency (JAEA), Tokai 319-1195, Japan}
\affiliation{Nishina Center for Accelerator-Based Science, RIKEN, Wako 351-0198, Japan}

\author{Kei~Suzuki}
\email[]{{\tt k.suzuki.2010@th.phys.titech.ac.jp}}
\affiliation{Advanced Science Research Center, Japan Atomic Energy Agency (JAEA), Tokai 319-1195, Japan}

\date{\today}

\begin{abstract}
The mass spectra of singly charmed and bottom baryons, $\Lambda_{c/b}(1/2^\pm,3/2^-)$ and $\Xi_{c/b}(1/2^\pm,3/2^-)$, are investigated using a nonrelativistic potential model with a heavy quark and a light diquark.
The masses of the scalar and pseudoscalar diquarks are taken from a chiral effective theory.
The effect of $U_A(1)$ anomaly induces an inverse hierarchy between the masses of strange and nonstrange pseudoscalar diquarks, which leads to a similar inverse mass ordering in $\rho$-mode excitations of singly heavy baryons.
\end{abstract}


\maketitle

\section{Introduction}
Diquarks, strongly correlated two-quark states, have a long history in hadron physics since the 1960s \cite{GellMann:1964nj,Ida:1966ev,Lichtenberg:1967zz,Lichtenberg:1967,Souza:1967rms,Lichtenberg:1968zz,Carroll:1969ty,Lichtenberg:1969pp} (see Refs.~\cite{Anselmino:1992vg,Jaffe:2004ph} for reviews).
It is an important concept for understanding the various physics in quantum chromodynamics (QCD), such as the baryon (and also exotic-hadron) spectra as well as the color superconducting phase.
The properties of various diquarks, such as the mass and size, have been studied by lattice QCD simulations~\cite{Hess:1998sd,Orginos:2005vr,Alexandrou:2006cq,Babich:2007ah,DeGrand:2007vu,Green:2010vc,Bi:2015ifa}.

A phenomenon related to diquark degrees of freedom is the spectrum of singly heavy baryons ($Qqq$), where a baryon contains two light (up, down, or strange) quarks ($q=u,d,s$) and one heavy (charm or bottom) quark ($Q=c,b$), so that the two light quarks ($qq$) might be well approximated as a diquark (for model studies about diquarks in $Qqq$ baryons, {\it e.g.}, see Refs.~\cite{Lichtenberg:1975ap,Lichtenberg:1982jp,Lichtenberg:1982nm,Fleck:1988vm,Ebert:1995fp,Ebert:2007nw,Kim:2011ut,Ebert:2011kk,Chen:2014nyo,Jido:2016yuv,Kumakawa:2017ffl,Harada:2019udr}).
In particular, the spectrum of singly heavy baryons is a promising candidate visibly affected by diquark degrees of freedom.
For example, the $P$-wave excited states of singly heavy baryons are classified by {\it $\lambda$ modes} (the orbital excitations between the diquark and heavy quark) and {\it $\rho$ modes} (the orbital excitations between two light quarks inside the diquarks) \cite{Copley:1979wj,Yoshida:2015tia}.

The chiral symmetry and $U_A(1)$ symmetry are fundamental properties of light quarks in QCD, and in the low-energy region of QCD they are broken by the chiral condensates and $U_A(1)$ anomaly, respectively.
Such symmetry breaking effects should be related to the properties of diquarks~\cite{Hong:2004xn,Hong:2004ux,Harada:2019udr}.
In Ref.~\cite{Harada:2019udr}, a chiral effective theory based on the $SU(3)_R \times SU(3)_L$ chiral symmetry with the scalar ($J^P = 0^+$, where $J$ and $P$ are the total angular momentum and parity, respectively) diquarks belonging to the color antitriplet $\bf \bar{3}$ and flavor antitriplet $\bf \bar{3}$ channel and its pseudoscalar ($0^-$) counterpart was constructed.\footnote{The diquark with the color $\bf \bar{3}$ and flavor $\bf \bar{3}$ is often referred to as the ``good" diquark~\cite{Jaffe:2004ph,Wilczek:2004im}.} 
These are the following new (and interesting) suggestions:
\begin{enumerate}
\item[(i)] {\it Chiral partner structures of diquarks}---A scalar diquark and its pseudoscalar partner belong to a chiral multiplet, which is the so-called chiral partner structure.
This structure means that chiral partners are degenerate when the chiral symmetry is completely restored.
As a result, they also predicted a similar chiral partner structure for charmed baryons such as $\Lambda_c(1/2^+)$-$\Lambda_c(1/2^-)$ and $\Xi_c(1/2^+)$-$\Xi_c(1/2^-)$ (for similar studies, see Refs.~\cite{Kawakami:2018olq,Kawakami:2019hpp}).
\item[(ii)] {\it Inverse hierarchy of diquark masses}---The effect of the $U_A(1)$ anomaly leads to an inverse hierarchy for the masses of the pseudoscalar diquarks: $M(us/ds, 0^-)<M(ud,0^-)$.
This is contrary to an intuitive ordering $M(ud,0^-)<M(us/ds, 0^-)$ expected from the larger constituent mass of the $s$ quark than that of the $u$ and $d$ quarks.
As a result of the inverse hierarchy, they also predicted a similar ordering for the charmed baryons: $M(\Xi_c,1/2^-)<M(\Lambda_c(1/2^-))$.
\end{enumerate}

In this paper, we investigate the spectrum of singly heavy baryons by using a ``hybrid" approach with the constituent diquarks based on the chiral effective theory~\cite{Harada:2019udr} and nonrelativistic two-body potential model (sometimes simply called {\it quark-diquark model}).
Our approach has the following advantages:
\begin{enumerate}
\item[(i)]
It can study the singly heavy-baryon spectrum based on the chiral partner structures of diquarks.
\item[(ii)]
It can introduce the inverse hierarchy of the pseudoscalar diquark masses originated from the $U_A(1)$ anomaly and examine its effects on the singly heavy baryons.
\item[(iii)]
It can take into account the contribution from the confining (linear and Coulomb) potential.
This is an additional advantage missing in Ref.~\cite{Harada:2019udr}.
\item[(iv)]
It can predict $\lambda$-mode excited states of singly heavy baryons.
This is more profitable than the approach in Ref.~\cite{Harada:2019udr}, where it will be difficult to calculate $\lambda$-mode excitations only by the effective Lagrangian though the $\rho$-mode states are naively estimated.
\end{enumerate}

It should be noted that the diquark-heavy-quark approach can cover all the excitation modes that appear in the conventional quark model. The orbital excitations of a three-quark system consist of the $\lambda$-mode, in which the diquark is intact, and the $\rho$-mode, in which the diquark is internally excited. The latter can be represented by a new type of diquark.
In the present approach, we consider only the scalar and pseudoscalar diquarks, but there are many other possible diquarks~\cite{Harada:2019udr}. Among them the vector and axial-vector diquarks are known to be low-lying and play major roles in the flavor {\bf 6} baryons, such as $\Sigma_c$ and $\Omega_c$. The chiral effective theory for the vector and axial-vector diquarks and their couplings to the scalar and pseudoscalar diquarks is being considered in the forthcoming paper.

This paper is organized as follows.
In Sec.~\ref{Sec_2}, we formulate the hybrid approach of the chiral effective theory and the potential model.
In Sec.~\ref{Sec_3}, we show the numerical results.
Section~\ref{Sec_4} is devoted to our conclusion and outlook.

\section{Formalism} \label{Sec_2}
In this section, we summarize the mass formulas of diquarks based on the chiral effective theory~\cite{Harada:2019udr}.
After that, we construct a nonrelativistic potential model for singly heavy baryons composed of a heavy quark and a diquark.

\subsection{Chiral effective Lagrangian}
In this work, we concentrate on the scalar ($0^+$) and pseudoscalar ($0^-$) diquarks with color $\bf \bar{3}$ and flavor $ \bf \bar{3}$. 
In the chiral effective theory of diquarks \cite{Harada:2019udr}, we consider the right-handed and left-handed diquark fields, $d_{R,i}$  and $d_{L,i}$,
where $i$ is the flavor index of a diquark.
The $i=1$ ($ds$) and $i=2$ ($su$) diquarks include one strange quark, while the $i=3$ ($ud$) diquark has no strange quark.

When the chiral symmetry and flavor $SU(3)$ symmetry are broken, the mass terms for the diquarks are given by~\cite{Harada:2019udr} 
\begin{eqnarray}
 {\cal L}_{\rm mass}
&=& - m_{0}^2 (d_{R,i} d_{R,i}^\dagger  +d_{L,i} d_{L,i}^\dagger)\nonumber\\
&&  - (m_{1}^2+A m_2^2 ) (d_{R,1} d_{L,1}^\dagger +d_{L,1} d_{R,1}^\dagger
\nonumber\\
&& +d_{R,2} d_{L,2}^\dagger +d_{L,2} d_{R,2}^\dagger) \nonumber\\
&& - (Am_{1}^2+m_2^2 ) ( d_{R,3} d_{L,3}^\dagger +d_{L,3} d_{R,3}^\dagger),
\label{SBLagrangian-mass}
\end{eqnarray}
where $m_0$, $m_1$, and $m_2$ are the model parameters.
$m_0$ is called the {\it chiral invariant mass}.
The term with $m_0^2$ satisfies the chiral symmetry, while the terms with $m_1^2$ and $m_2^2$ break the chiral symmetry spontaneously and explicitly.
$m_1$ and $m_2$ are the coefficients of the six-point quark (or diquark-duquark-meson) interaction motivated by the $U_A(1)$ anomaly and the eight-point quark (or diquark-duquark-meson-meson) interaction which conserves the $U_A(1)$ symmetry, respectively.
$A\sim 5/3$ is the parameter of the flavor $SU(3)$ symmetry breaking due to the quark mass difference, $m_s > m_u \simeq m_d$.

\subsection{Mass formulas of diquarks}
By diagonalizing the mass matrix (\ref{SBLagrangian-mass}) in $R/L$ and flavor space, we obtain the mass formulas for the diquarks, $M_i(0^\pm)$ \cite{Harada:2019udr}:
\begin{eqnarray}
&& M_1(0^+) = M_2(0^+) = \sqrt{m_0^2 - m_{1}^2 - Am_2^2} ,\label{M1P}\\
&& M_3(0^+) =  \sqrt{m_0^2 - Am_{1}^2 - m_2^2}, \label{M3P}\\
&& M_1(0^-) = M_2(0^-) = \sqrt{m_0^2 +m_{1}^2 +Am_2^2}, \label{M1M}\\
&& M_3(0^-) = \sqrt{m_0^2 + Am_{1}^2 +m_2^2} .\label{M3M}
\end{eqnarray}
From Eqs.~(\ref{M1P})-(\ref{M3M}), we get
\begin{eqnarray}
&& \left[ M_{1,2}(0^+) \right]^2 - \left[ M_{3}(0^+) \right]^2 
= \left[ M_{3}(0^-) \right]^2 - \left[ M_{1,2}(0^-) \right]^2 \nonumber\\
&&\qquad= ( A - 1) ( m_1^2 - m_2^2 ).
\label{rel-diff}
\end{eqnarray}
From this relation with $A>1$ and $m_1^2>m_2^2$, one finds the inverse mass hierarchy for the pseudoscalar diquarks: $M_{3}(0^-) > M_{1,2}(0^-) $, where the nonstrange diquark ($i=3$) is heavier than the strange diquark ($i=1,2$).

\subsection{Potential quark-diquark model}
In order to calculate the spectrum of singly heavy baryons, we apply a nonrelativistic two-body potential model with a single heavy quark and a diquark.

The nonrelativistic two-body Hamiltonian is written as
\begin{equation}
H =  \frac{\bm{p}_Q^2}{2M_Q} + \frac{\bm{p}_d^2}{2M_d} + M_Q + M_d + V(r),
\end{equation}
where the indices $Q$ and $d$ denote the heavy quark and diquark, respectively.
$\bm{p}_{Q/d}$ and $M_{Q/d}$ are the momentum and mass, respectively.
$r=\bm{r}_d-\bm{r}_Q$ is the relative coordinate between the two particles.
After subtracting the kinetic energy of the center of mass motion, the Hamiltonian is reduced to
\begin{equation}
H =  \frac{\bm{p}^2}{2\mu} + M_Q + M_d + V(r),
\end{equation}
where $\bm{p} = \frac{M_Q \bm{p}_d - M_d \bm{p}_Q}{M_d+M_Q}$ and $\mu=\frac{M_dM_Q}{M_d+M_Q}$ are the relative momentum and reduced mass, respectively.

For the potential $V(r)$, in this work, we apply three types of potentials constructed by Yoshida {\it et al.}~\cite{Yoshida:2015tia}, Silvestre-Brac~\cite{SilvestreBrac:1996bg}, and Barnes {\it et al.}~\cite{Barnes:2005pb}.
These potentials consist of the Coulomb term with the coefficient $\alpha$ and the linear term with $\lambda$,
\begin{equation}
V(r) = - \frac{\alpha}{r}+\lambda r+C,
\end{equation}
where $C$ in the last term is a ``constant shift" of the potential, which is a model parameter depending on the specific system.

\begin{table*}[tbh!]
  \centering
  \caption{Potential model parameters used in this work.
We apply three types of potentials, Yoshida (Potential Y)~\cite{Yoshida:2015tia}, Silvestre-Brac (Potential S)~\cite{SilvestreBrac:1996bg}, and Barnes (Potential B)~\cite{Barnes:2005pb}.
$\alpha$, $\lambda$, $C_c$, $C_b$, $M_c$, and $M_b$ are the coefficients of Coulomb and linear terms, constant shifts for charmed and bottom baryons, and masses of constituent charm and bottom quarks, respectively.
$\mu$ is reduced mass of two-body systems.
The values of $C_c$ and $C_b$ are fitted by our model.
For Potential B, $M_b$ is not given~\cite{Barnes:2005pb}, so that we do not fit $C_b$.}
  \begin{tabular}{ l c c c c c c }
  \hline \hline
              & $\alpha$ & $ \lambda(\rm GeV^2)$  & $ C_c(\rm GeV)$ & $ C_b(\rm GeV)$ & $M_c(\rm GeV)$ & $M_b(\rm GeV)$ \\ \hline
    Potential Y~\cite{Yoshida:2015tia}   & $(2/3) \times 90(\rm MeV)/\mu$ & $0.165$&$-0.831$  & $-0.819$ & $1.750$ & $5.112$ \\ 
    Potential S~\cite{SilvestreBrac:1996bg} & $0.5069$ & $0.1653$&$-0.707$ & $-0.696$ & $1.836$ & $5.227$ \\ 
    Potential B~\cite{Barnes:2005pb}    & $(4/3) \times 0.5461$ & $0.1425$&$-0.191$ & $\ldots$ & $1.4794$ & $\ldots$ \\ \hline \hline
  \end{tabular}
  \label{tab:para}
\end{table*}

Note that, only in Ref.~\cite{Yoshida:2015tia}, the coefficient $\alpha$ of the Coulomb term depends on $1/\mu$.
In other word, this is a ``mass-dependent" Coulomb interaction, which is motivated by the behavior of the potential obtained from lattice QCD simulations \cite{Kawanai:2011xb}.
On the other hand, the other potentials \cite{SilvestreBrac:1996bg,Barnes:2005pb} do not include such an effect.
Such a difference between the potentials will lead to a quantitative difference also in singly heavy-baryon spectra.

In this work, the charm quark mass $M_c$, bottom quark mass $M_b$, $\alpha$, and $\lambda$ are fixed by the values estimated in the previous studies \cite{Yoshida:2015tia,SilvestreBrac:1996bg,Barnes:2005pb}, which are summarized in Table~\ref{tab:para}.
The other parameters are determined in Sec.~\ref{Sec_3-1}.

In order to numerically solve the Schr\"odinger equation, we apply the Gaussian expansion method \cite{Kamimura:1988zz,Hiyama:2003cu}.

\section{Numerical results} \label{Sec_3}
\subsection{Parameter determination} \label{Sec_3-1}

In this section, we determine the unknown model parameters such as the diquark masses $M_i(0^\pm)$ and constant shifts, $C_c$ for charmed baryons and $C_b$ for bottom baryons.
The procedure is as follows:
\begin{enumerate}
\item[(i)] {\it Determination of $C_c$}---By inputting the mass of the $ud$ scalar diquark, $M_3 (0^+)$, we determine the constant shift $C_c$ so as to reproduce the ground-state mass of $\Lambda_c$.
As $M_3 (0^+)$, we apply the value measured from recent lattice QCD simulations with $2+1$ dynamical quarks~\cite{Bi:2015ifa}: $M_3 (0^+)=725 \ \rm{MeV}$.
As the mass of $\Lambda_c$, we use the experimental value from PDG~\cite{Tanabashi:2018oca}: $M(\Lambda_c,1/2^+)=2286.46 \ \rm{MeV}$.

\item[(ii)] {\it Determination of $ M_{1,2} (0^+)$ and $ M_{3} (0^-)$}---After fixing $C_c$, we next fix two diquark masses, $ M_{1,2} (0^+)$ and $ M_{3} (0^-)$.
Here, we apply the following two methods:
\begin{enumerate}
\item[] {\it Model I.}---The first method is to input $ M_{1,2} (0^+)$ and $ M_{3} (0^-)$ measured from recent lattice QCD simulations~\cite{Bi:2015ifa}: $M_{1,2} (0^+)=906 \ \rm{MeV}$ and $M_{3} (0^-)=1265 \ \rm{MeV}$.\footnote{We use $ M_{3} (0^+)$ and $ M_{1,2} (0^+)$ in the chiral limit in Ref.~\cite{Bi:2015ifa}.
The chiral extrapolation of $M_3(0^-)$ is not shown in Ref.~\cite{Bi:2015ifa}, so that we use $M_3(0^-)$ at the lowest quark mass (see Table 8 of Ref.~\cite{Bi:2015ifa}).}
We call the choice of these parameters {\it Model I}, which is similar to Method I in Ref.~\cite{Harada:2019udr}.

\item[]{\it Model II.}---Another method is to determine $ M_{1,2} (0^+)$ and $ M_{3} (0^-)$ from the potential model and some known baryon masses, which we call {\it Model II}.
After fixing $C_c$, $ M_{1,2} (0^+)$ and $ M_{3} (0^-)$ are determined so as to reproduce $M(\Xi_c,1/2^+)$ and $M_\rho(\Lambda_c,1/2^-)$, respectively.
As input parameters, we use the experimental values of the ground-state $\Xi_c$ from PDG~\cite{Tanabashi:2018oca}: $M(\Xi_c,1/2^+)=2469.42 \ \rm{MeV}$.
For the mass of the $\rho$ mode of the negative-parity $\Lambda_c$, we use the value predicted by a nonrelativistic three-body calculation in Ref.~\cite{Yoshida:2015tia}: $M_\rho(\Lambda_c,1/2^-)=2890 \ \rm{MeV}$.\footnote{Note that while the known experimental value of negative-parity $\Lambda_c$, $M(\Lambda_c,1/2^-) = 2592.25 \ \rm{MeV}$~\cite{Tanabashi:2018oca}, is expected to be that of the $\lambda$-mode excitation, the resonance corresponding to the $\rho$-mode has not been observed.}
\end{enumerate}

\item[(iii)] {\it Determination of $ M_{1,2} (0^-)$}---Using the mass relation~(\ref{rel-diff}) and our three diquark masses, $M_{3} (0^+)$, $M_{1,2} (0^+)$, and $ M_{1,2} (0^-)$, we determine the masses of $us/ds$ pseudoscalar diquarks, $M_{1,2} (0^-)$.
Here, we emphasize that the estimated $M_{1,2} (0^-)$ reflects the inverse hierarchy for the diquark masses, which has never been considered in previous studies except for Ref.~\cite{Harada:2019udr}.

\item[(iv)] {\it Determination of $C_b$}---For singly bottom baryons, we also determine the constant shift $C_b$ by inputting $M_3 (0^+)=725 \ \rm{MeV}$ and reproducing the mass of the ground state $\Lambda_b$, $M(\Lambda_b,1/2^+)=5619.60 \ \rm{MeV}$~\cite{Tanabashi:2018oca}.
For the diquark masses, $M_{1,2} (0^+)$, $M_{3} (0^-)$, and $M_{1,2} (0^-)$, we use the same values as the case of the charmed baryons.
\end{enumerate}

\begin{table*}[tbh!]
  \centering
  \caption{List of numerical values of scalar [$M_3 (0^+)$, $M_{1,2} (0^+)$] and pseudoscalar [$M_3 (0^-)$, $M_{1,2} (0^-)$] diquark masses, masses of singly-heavy baryons [$M(\Lambda_Q)$, $M(\Xi_Q)$], coefficients of chiral effective Lagrangian ($m_0^2$, $m_1^2$, and $m_2^2$).
We compare the results from our approach using three potential and two parameters, Yoshida (denoted as IY and IIY), Silvestre-Brac (IS and IIS), and Barnes (BI and IIS) with a naive estimate in the chiral EFT \cite{Harada:2019udr} (Method~I and Method~II) and the experimental values from PDG~\cite{Tanabashi:2018oca}.
The asterisk ($*$) denotes the input values.} 
  \begin{tabular}{ l c c c c c c c c c} 
   \hline \hline     
        \multicolumn{1}{ l }{}
      & \multicolumn{2}{c}{Chiral EFT~\cite{Harada:2019udr}} 
      & \multicolumn{6}{c}{Potential model (this work)}
      & \multicolumn{1}{c}{} \\ \cline{2 - 9} 
Mass $(\rm MeV)$                           &Method~I &Method~II& IY & IS & IB & IIY & IIS & IIB & Experiment~\cite{Tanabashi:2018oca} \\
\hline \hline
    $ M_3 (0^+)$                        & $725^*$ & $725^*$ & $725^*$ & $725^*$ & $725^*$ & $725^*$& $725^*$& $725^*$& \\
    $ M_{1,2} (0^+) $                   & $906^*$ & $906$   & $906^*$ & $906^*$ & $906^*$ & $942$  & $977$  & $983$  & \\
    $ M_3 (0^-) $                       & $1265^*$& $1329$  & $1265^*$&$1265^*$ &$1265^*$ & $1406$ & $1484$ & $1496$ & \\ 
    $ M_{1,2} (0^-) $                   & $1142$  & $1212$  & $1142$  & $1142$  & $1142$  & $1271$ & $1331$ & $1341$ & \\
\hline    
    $ M(\Lambda_c,1/2^+)$               & $2286^*$&$2286^*$ & $2286^*$ & $2286^*$& $2286^*$& $2286^*$ & $2286^*$ & $2286^*$ & $ 2286.46 $ \\
    $ M(\Xi_c,1/2^+)$                   & $2467$  &$2469^*$ & $2438$ & $2415$ & $2412$ & $2469^*$ & $2469^*$ & $2469^*$ & $ 2469.42 $ \\
    $ M_\rho(\Lambda_c,1/2^-)$          & $2826$  &$2890^*$ & $2759$ & $2702$ & $2694$ & $2890^*$ & $2890^*$ & $2890^*$ & $\ldots$ \\
    $ M_\rho(\Xi_c,1/2^-)$              & $2704$  &$2775$   & $2647$ & $2600$ & $2594$ & $2765$   & $2758$   & $2758$   & $(2793.25)$ \\ 
    $ M_\lambda(\Lambda_c,1/2^-,3/2^-)$&$\ldots$  &$\ldots$ & $2613$ & $2703$ & $2734$ & $2613$   & $2703$   & $2734$   & $(2616.16)$ \\
    $ M_\lambda(\Xi_c,1/2^-,3/2^-)$    &$\ldots$  &$\ldots$ & $2748$ & $2825$ & $2860$ & $2776$   & $2878$   & $2918$   & $(2810.05)$ \\
\hline         
    $ M(\Lambda_b,1/2^+)$               &$\ldots$ &$\ldots$ & $5620$ & $5620$ &$\ldots$ & $5620^*$ & $5620^*$ &$\ldots$  & $5619.60$ \\
    $ M(\Xi_b,1/2^+)$                   &$\ldots$ &$\ldots$ & $5766$ & $5735$ &$\ldots$ & $5796$   & $5785$   &$\ldots$  & $5794.45$ \\
    $ M_\rho(\Lambda_b,1/2^-)$          &$\ldots$ &$\ldots$ & $6079$ & $5999$ &$\ldots$ & $6207$   & $6174$   &$\ldots$  & $(5912.20)$ \\
    $ M_\rho(\Xi_b,1/2^-)$              &$\ldots$ &$\ldots$ & $5970$ & $5905$ &$\ldots$ & $6084$   & $6051$   &$\ldots$  & $\ldots$ \\ 
    $ M_\lambda(\Lambda_b,1/2^-,3/2^-)$ &$\ldots$ &$\ldots$ & $5923$ & $6028$ &$\ldots$ & $5923$   & $6028$   &$\ldots$  & $(5917.35)$ \\
    $ M_\lambda(\Xi_b,1/2^-,3/2^-)$     &$\ldots$ &$\ldots$ & $6049$ & $6139$ &$\ldots$ & $6076$   & $6188$   &$\ldots$  & $\ldots$ \\
\hline
Parameter $(\rm MeV^2)$                 & & & & & & & & & \\
\hline
    $m_0^2$                    &$(1031)^2$ &$(1070)^2$ & $(1031)^2$ & $(1031)^2$ & $(1031)^2$ & $(1119)^2$ & $(1168)^2$ & $(1176)^2$ & \\
    $m_1^2$                    &$(606)^2$  &$(632)^2$  & $(606)^2$  & $(606)^2$ & $(606)^2$ & $(690)^2$  & $(746)^2$  & $(754)^2$  & \\
    $m_2^2$                    &$-(274)^2$ &$-(213)^2$ & $-(274)^2$ & $-(274)^2$ & $-(274)^2$ & $-(258)^2$ & $-(298)^2$ & $-(303)^2$ & \\
\hline \hline   
  \end{tabular}
  \label{tab:spectrum}
\end{table*}

The constant shifts, $C_c$ and $C_b$, estimated by us are summarized in Table~\ref{tab:para}.
The diquark masses predicted by us are shown in Table~\ref{tab:spectrum}.
By definition, the diquark masses in Model I are the same as the values from Method I in Ref.~\cite{Harada:2019udr}.
Here we focus on the comparison of the prediction from Model IIY and that from Method II in Ref.~\cite{Harada:2019udr}.
In both the approaches, the input values of $M(\Xi_c,1/2^+) = 2469 \ \rm MeV$ and $ M_\rho(\Lambda_c,1/2^-) = 2890 \ \rm MeV$ are the same.
Our prediction is $M_{1,2} (0^+) = 942 \ \rm MeV$, which is larger than $906 \ \rm MeV$ estimated in Ref.~\cite{Harada:2019udr}.
This difference is caused by the existence of the confining potential (particularly, linear potential) which is not considered in the estimate in Ref.~\cite{Harada:2019udr}.
This tendency does not change in the results using the other potentials.
Similarly, for $M_{3} (0^-)$, we obtain $1406 \ \rm MeV$, which is significantly larger than $1329 \ \rm MeV$ in Ref.~\cite{Harada:2019udr}.

Next, we focus on the ordering of the pseudoscalar diquarks.
We find the inverse hierarchy $M_{1,2} (0^-) < M_{3} (0^-)$ in all the models, which is consistent with the prediction in Ref.~\cite{Harada:2019udr}.
We emphasize that the inverse mass hierarchy of diquarks does not suffer from the confining potential.

Furthermore, from the diquark masses and Eqs.~(\ref{M1P})--(\ref{M3M}), we can determine the unknown parameters of chiral effective Lagrangian, $m_0$, $m_1$, and $m_2$, which is also summarized in Table~\ref{tab:spectrum}.
By definition, the values in Model I are the same as those from Method I in Ref.~\cite{Harada:2019udr}.
Here we compare our estimate from Model II and a naive estimate from Method II in Ref.~\cite{Harada:2019udr}.
From Models IIY, IIS, and IIB, we conclude that these parameters are insensitive to the choices of the quark model potential.
We also see that inclusion of the confining potential does not alter the parameters qualitatively.
Quantitatively, the magnitude of these parameters is larger than that from Method II in Ref.~\cite{Harada:2019udr}, which is expected to be improved by taking into account the confining potential.

\subsection{Spectrum of singly charmed baryons}
\begin{figure*}[tbh!]
    \begin{center}
            \includegraphics[clip,width=2.0\columnwidth]{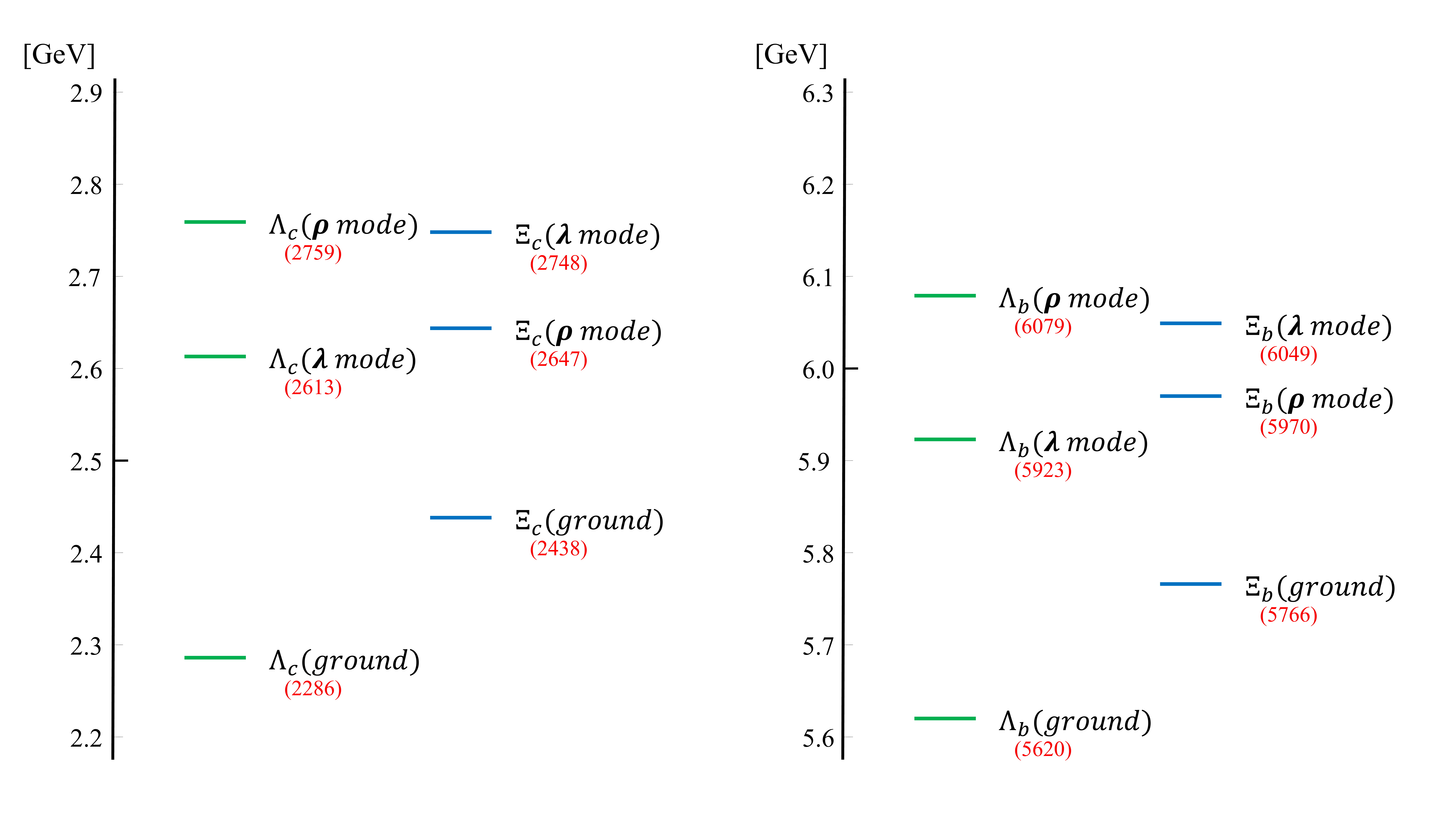}
    \end{center}
    \caption{The energy spectra of singly charmed and bottom baryons from our numerical results using Model IY.}
    \label{fig:spectrum1}
\end{figure*}

\begin{figure*}[tbh!]
    \begin{center}
            \includegraphics[clip,width=2.0\columnwidth]{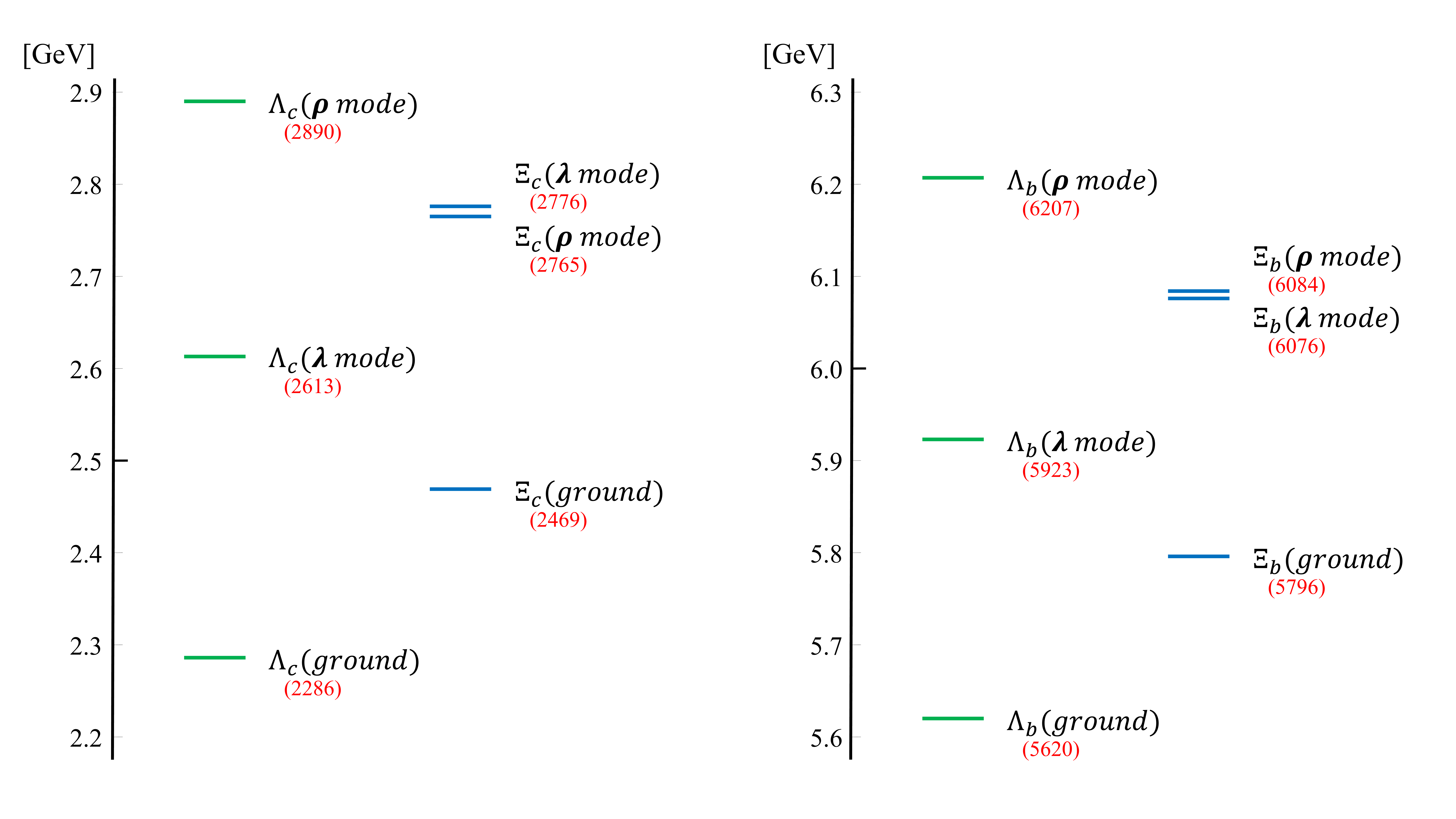}
    \end{center}
    \caption{The energy spectra of singly charmed and bottom baryons from our numerical results using Model IIY.}
    \label{fig:spectrum2}
\end{figure*}

The values of masses of singly charmed baryons are summarized in Table~\ref{tab:spectrum}.
$ M(\Lambda_c,1/2^+)$, $M(\Xi_c,1/2^+)$, and $M_\rho(\Lambda_c,1/2^-)$ are the input values.
Similarly to the ordering of $M_{1,2} (0^-) < M_{3} (0^-)$, we find the inverse hierarchy for the $\rho$-mode excitations of the singly charmed baryons: $M_\rho(\Xi_c,1/2^-) < M_\rho(\Lambda_c,1/2^-)$.
This is our main conclusion: {\it the inverse mass hierarchy between the $\rho$ mode of $\Lambda_c$ (without a strange quark) and that of $\Xi_c$ (with a strange quark) is realized even with the confining potential}, which is consistent with the naive estimate with the chiral effective theory~\cite{Harada:2019udr}.

The energy spectra for $\Lambda_c$ and $\Xi_c$ from Models IY and IIY are shown in the left panels of Figs.~\ref{fig:spectrum1} and \ref{fig:spectrum2}.
Here we emphasize the qualitative difference between the spectra of the negative-parity $\Lambda_c$ and $\Xi_c$.
In the $\Lambda_c$ spectrum, the $\rho$ mode is heavier than the $\lambda$ mode, which is consistent with the three-body calculation~\cite{Yoshida:2015tia}.
On the other hand, in the $\Xi_c$ spectrum, the $\rho$ and $\lambda$ modes are close to each other.
As a result, the mass splitting between the $\rho$ and $\lambda$ modes in the $\Xi_c$ spectrum is smaller than that in the $\Lambda_c$ spectrum: $|M_\rho(\Xi_c,1/2^-) - M_\lambda(\Xi_c,1/2^-,3/2^-)| < |M_\rho (\Lambda_c,1/2^-) - M_\lambda(\Lambda_c,1/2^-,3/2^-)|$, where we note that the $1/2^-$ and $3/2^-$ states for $\lambda$ modes in our model are degenerate as discussed later.
The significant difference between Models IY and IIY is caused by $M_\rho (\Lambda_c,1/2^-)$ which is related to $M_3(0^-)$.
From the diquark mass relation~(\ref{rel-diff}), a larger $M_3(0^-)$ leads to a larger $M_{1,2}(0^-)$.
Then a heavier $M_\rho (\Lambda_c,1/2^-)$ leads to a heavier $M_\rho (\Xi_c,1/2^-)$.
As a result, $M_\rho (\Xi_c,1/2^-)$ from Model IIY is heavier than $M_\rho (\Xi_c,1/2^-)$ from Model IY. 

Next, we discuss the masses of the $\lambda$ modes.
The $\lambda$ modes are the excited states with the orbital angular momentum between the heavy quark and diquark, so that their masses are higher than those of the ground states, which is the ``$P$-wave" states in our two-body potential model.
Also, in singly heavy-baryon spectra, the masses of the $\lambda$ modes are usually lower than that of the $\rho$ modes, as shown by the three-body calculation~\cite{Yoshida:2015tia}. 
In Models IY and IIY, the excitation energy from the ground state, $M_\lambda(\Lambda_c,1/2^-,1/3^-) - M(\Lambda_c,1/2^+)$, is about $300 \ \rm{MeV}$.
For the other potentials, it is more than $400 \ \mathrm{MeV}$. 
This difference is caused by the coefficients $\alpha$ of the Coulomb interaction.
In the Yoshida potential used in Models IY and IIY, $\alpha$ is relatively small, so that its wave function is broader.
As a result, the difference between the wave functions of the ground and excited states becomes smaller, and the excitation energy also decreases.

The known experimental values of the negative-parity $\Lambda_c$ and $\Xi_c$ are $M(\Lambda_c,1/2^-) = 2592.25 \ \rm{MeV}$, $M(\Lambda_c,3/2^-) = 2628.11 \ \rm{MeV}$, $M(\Xi_c,1/2^-) = 2793.25 \ \rm{MeV}$, and $M(\Xi_c,3/2^-) = 2818.45 \ \rm{MeV}$~\cite{Tanabashi:2018oca}.
The $\lambda$ modes in our results correspond to the spin average of $1/2^-$ and $3/2^-$.
The spin averages of the experimental values are $M(\Lambda_c,1/2^-,3/2^-) = 2616.16 \ \rm{MeV}$ and $M(\Xi_c,1/2^-,3/2^-) = 2810.05 \ \rm{MeV}$.
For the negative-parity $\Lambda_c$, the experimental value of $M(\Lambda_c,1/2^-,3/2^-)$ is expected to be $\lambda$ modes.
Then our predictions from Models IY and IIY are in good agreement with the experimental value.
If the experimental value of $M(\Lambda_c,1/2^-)$ is assigned to the $\rho$ mode, it is much smaller than our prediction.
For the negative-parity $\Xi_c$, when the experimental value of $M(\Xi_c,1/2^-,3/2^-)$ is assigned to the $\lambda$ modes, the value is close to our results from Models IS, IB, and IIY within $50 \ \rm{MeV}$.
When the experimental value of $M(\Xi_c,1/2^-)$ is assigned to the $\rho$ mode, the value is close to our results from Models IIY, IIS, and IIB within $50 \ \rm{MeV}$.
Thus, Model IIY can reproduce the known experimental values in any case.
In addition, when these experimental values are assigned to the $\lambda$ modes, the excitation energy of the $\lambda$ modes from the ground state is estimated to be about $330$-$340 \ \rm{MeV}$, which is consistent with the results from Models IY and IIY.

We comment on the possible splitting in the $\lambda$ modes.
The splitting between $1/2^-$ and $3/2^-$ states is caused by the spin-orbit (LS) coupling.
In order to study this splitting within our model, we need to introduce the LS coupling between the orbital angular momentum and the heavy-quark spin.
In the heavy-quark limit ($m_c \to \infty$), the two states are degenerate due to the suppression of the LS coupling, so that they are called the {\it heavy-quark spin doublet}.

\subsection{Spectrum of singly bottom baryons}
For the singly bottom baryons, the input value is only the mass of the ground-state $\Lambda_b(1/2^+)$, and here we give predictions for the other states.
For the ground state of $\Xi_b(1/2^+)$, our prediction with Models IY and IIY is in good agreement with the known mass $M(\Xi_b,1/2^+) = 5794.45 \ \rm{MeV}$~\cite{Tanabashi:2018oca}.
This indicates that the quark-diquark picture is approximately good for $\Xi_b(1/2^+)$.

The energy spectra for $\Lambda_b$ and $\Xi_b$ from Models IY and IIY are shown in the right panels of Figs.~\ref{fig:spectrum1} and \ref{fig:spectrum2}.
Similarly to the charmed baryon spectra, we again emphasize the difference between the $\Lambda_b$ and $\Xi_b$ spectra.
For the $\rho$ modes, we also find the inverse mass hierarchy: $M_\rho(\Xi_b,1/2^-) < M_\rho(\Lambda_b,1/2^-)$.
The difference between Models IY and IIY is similar to the charmed baryons.

The known experimental values of negative-parity $\Lambda_b$ are $M(\Lambda_b,1/2^-) = 5912.20 \ \rm{MeV}$ and $M(\Lambda_b,3/2^-) = 5919.92 \ \rm{MeV}$~\cite{Tanabashi:2018oca}, and their spin average is $M(\Lambda_b,1/2^-,3/2^-) =5917.35 \ \rm{MeV}$.
Whether these states are the $\rho$ mode or $\lambda$ mode is not determined yet.
When the experimental value of $M(\Lambda_b,1/2^-,3/2^-)$ is assigned to the $\lambda$ modes, the value is in agreement with the results from Models IY and IIY.
On the other hand, when the experimental value of $M(\Lambda_b,1/2^-)$ is assigned to the $\rho$ mode, it is quite smaller than our prediction.
This fact indicates that the experimental values correspond to $\lambda$ modes.
The negative-parity $\Xi_b$ is still not observed experimentally.
In 2018, a heavier state $\Xi_b(6227)$ with $M(\Xi_b)=6226.9 \ \rm{MeV}$ was observed~\cite{Tanabashi:2018oca,Aaij:2018yqz}, but its spin and parity are not determined so far.

\begin{table*}[tbh!]
  \centering
  \caption{Rms distance $\sqrt{\hat{r}^2}$ between a heavy quark and a diquark.}
  \begin{tabular}{ l c c c c c c} 
\hline \hline
Rms distance   $\rm (fm)$                          & \ \ \ IY \ \ \ & \ \ \ IS \ \ \ & \ \ \ IB \ \ \ & \ \ \ IIY \ \ \ & \ \ \ IIS \ \ \ & \ \ \ IIB \ \ \ \\
\hline
    $ \sqrt{\hat{r}^2} (\Lambda_c,1/2^+)$          & $0.587$ & $0.512$ & $0.506$ & $0.587$ & $0.512$ & $0.506$ \\
    $ \sqrt{\hat{r}^2} (\Xi_c,1/2^+)$              & $0.559$ & $0.476$ & $0.469$ & $0.555$ & $0.466$ & $0.457$ \\
    $ \sqrt{\hat{r}^2}_\rho (\Lambda_c,1/2^-)$     & $0.523$ & $0.431$ & $0.421$ & $0.513$ & $0.412$ & $0.402$ \\
    $ \sqrt{\hat{r}^2}_\rho (\Xi_c, 1/2^-)$        & $0.534$ & $0.444$ & $0.435$ & $0.523$ & $0.425$ & $0.415$ \\ 
    $ \sqrt{\hat{r}^2}_\lambda (\Lambda_c, 1/2^-,3/2^-)$ & $0.832$ & $0.783$ & $0.814$ & $0.832$ & $0.783$ & $0.814$ \\
    $ \sqrt{\hat{r}^2}_\lambda (\Xi_c, 1/2^-,3/2^-)$     & $0.792$ & $0.738$ & $0.767$ & $0.785$ & $0.724$ & $0.752$ \\
\hline  
    $ \sqrt{\hat{r}^2} (\Lambda_b, 1/2^+)$         & $0.548$ & $0.466$ & $\ldots$ & $0.548$ & $0.466$ & $\ldots$ \\
    $ \sqrt{\hat{r}^2} (\Xi_b, 1/2^+)$             & $0.515$ & $0.424$ & $\ldots$ & $0.510$ & $0.412$ & $\ldots$ \\
    $ \sqrt{\hat{r}^2}_\rho (\Lambda_b, 1/2^-)$    & $0.471$ & $0.368$ & $\ldots$ & $0.459$ & $0.346$ & $\ldots$ \\
    $ \sqrt{\hat{r}^2}_\rho (\Xi_b, 1/2^-)$        & $0.484$ & $0.384$ & $\ldots$ & $0.471$ & $0.362$ & $\ldots$ \\ 
    $ \sqrt{\hat{r}^2}_\lambda (\Lambda_b, 1/2^-,3/2^-)$ & $0.776$ & $0.724$ & $\ldots$ & $0.776$ & $0.724$ & $\ldots$ \\
    $ \sqrt{\hat{r}^2}_\lambda (\Xi_b, 1/2^-,3/2^-)$     & $0.728$& $0.671$ & $\ldots$ & $0.720$ & $0.654$ & $\ldots$ \\
\hline \hline
  \end{tabular}
  \label{tab:RMS}
\end{table*}

\subsection{Root-mean-square distance}
We summarize the root-mean-square (rms) distance, $\sqrt{\hat{r}^2}$, between the diquark and the heavy quark in Table~\ref{tab:RMS}.
We find the rms distance of the $\rho$ mode is smaller than those of the ground states and the $\lambda$ mode.
This is because the pseudoscalar diquark is heavier than the scalar diquark, $M(0^-)>M(0^+)$.
Then the kinetic energy of the system with $M(0^-)$ is suppressed, and, as a result, the wave function shrinks compared to its ground state with $M(0^+)$.
Due to the inverse hierarchy of the diquark masses, we find also the inverse hierarchy for the rms distance, $ \sqrt{\hat{r}^2}_\rho (\Lambda_c,1/2^-) < \sqrt{\hat{r}^2}_\rho (\Xi_c,1/2^-)$, which is different from the standard hierarchy seen in the ground states, $\sqrt{\hat{r}^2} (\Lambda_c,1/2^+) > \sqrt{\hat{r}^2} (\Xi_c,1/2^+)$.

The $\lambda$ modes are the $P$-wave excitations within a two-body quark-diquark model, so that their rms distance is larger than those of the ground and $\rho$-mode states which is ``$S$-wave" states within our model.
The rms distances in the bottom baryons are shorter than those of the charmed baryons because of the heavier bottom quark mass.

We find that the rms distances from Models IY and IIY are larger than those from the other models IS, IB, IIS, and IIB.
This difference is caused by the coefficient $\alpha$ of the attractive Coulomb interaction.
The Yoshida potential in Models IY and IIY has the relatively small $\alpha$, so that its wave function and the rms distance are larger than those from other models.

Note that the real wave function of a diquark must have a size which is the distance between a light quark and another light quark.
In our approach, namely, the quark-diquark model, diquarks are treated as a point particle, so that such a size effect is neglected.
To introduce such an effect would be important for improving our model.
In particular, it would be interesting to investigate the form factors of singly heavy baryons with the negative parity by lattice QCD simulations and to compare it with our predictions. 

\section{Conclusion and outlook} \label{Sec_4}
In this paper, we investigated the spectrum of singly heavy baryons using the hybrid approach of the chiral effective theory of diquarks and nonrelativistic quark-diquark potential model.

Our findings are as follows:
\begin{enumerate}
\item[(i)] We found the inverse mass hierarchy in the $\rho$-mode excitations of singly heavy baryons, $M(\Xi_Q,1/2^-)<M(\Lambda_Q,1/2^-)$, which is caused by the inverse mass hierarchy of the pseudoscalar diquarks $M(us/ds, 0^-)<M(ud,0^-)$.
This conclusion is the same as the naive estimate in Ref.~\cite{Harada:2019udr}, but it is important to note that the effect from the confining potential between a heavy quark and a diquark does not change this conclusion. 
\item[(ii)] We found that the mass splitting between the $\rho$- and $\lambda$-mode excitations in the $\Xi_Q$ spectrum is smaller than that in the $\Lambda_Q$ spectrum: $|M_\rho(\Xi_Q,1/2^-) - M_\lambda(\Xi_Q,1/2^-,3/2^-)| < |M_\rho (\Lambda_Q,1/2^-) - M_\lambda(\Lambda_Q,1/2^-,3/2^-)|$.
\end{enumerate}

The inverse mass hierarchy in singly heavy baryons can be also investigated by future lattice QCD simulations, as studied with quenched simulations \cite{Bowler:1996ws,AliKhan:1999yb,Woloshyn:2000fe,Lewis:2001iz,Mathur:2002ce,Flynn:2003vz,Chiu:2005zc}, as well as with dynamical quarks~\cite{Na:2007pv,Na:2008hz,Liu:2009jc,Alexandrou:2012xk,Briceno:2012wt,Namekawa:2013vu,Brown:2014ena,Bali:2015lka,Alexandrou:2017xwd,Can:2019wts,Bahtiyar:2020uuj}.
Although studying negative-parity baryons from lattice QCD is more difficult than the positive-parity states, there are a few works for singly heavy baryons~\cite{Chiu:2005zc,Bali:2015lka,Can:2019wts,Bahtiyar:2020uuj}.
Our findings give a motivation to examine the excited-state spectra from lattice QCD simulations.
Here, the careful treatment of the chiral and $U_A(1)$ symmetry on the lattice would be required.
Furthermore, the chiral effective Lagrangian for singly heavy baryons, as formulated in Sec. III-F of Ref.~\cite{Harada:2019udr}, is a useful approach for analytically studying the inverse mass hierarchy of heavy baryons.
In this Lagrangian, the assignment of the chiral partners for heavy baryons is the same as that for the diquarks in this work, so that we can obtain a similar spectrum.

The internal structures of excited states, such as $\rho$ and $\lambda$ modes, can significantly modify their decay properties~\cite{Chen:2007xf,Zhong:2007gp,Nagahiro:2016nsx,Chen:2017sci,Arifi:2017sac,Ye:2017yvl,Wang:2017kfr,Arifi:2018yhr,Guo:2019ytq}, and to study the decay processes taking into account the inverse hierarchy will be important.

In this paper, we focused only on the scalar diquark and its chiral partner.
As another important channel, the chiral-partner structure of the axial-vector ($1^+$) diquarks with the color $\bf \bar{3}$ and flavor $\bf 6$ (the so-called ``bad" diquarks~\cite{Jaffe:2004ph,Wilczek:2004im}) could be related to the spectra of $\Sigma_Q$, $\Sigma_Q^\ast$, $\Xi_Q^\prime$, $\Xi_Q^\ast$, $\Omega_Q$, and $\Omega_Q^\ast$ baryons.

Furthermore, the diquark correlations at high temperature are expected to modify the production rate of singly heavy baryons in high-energy collision experiments~\cite{Sateesh:1991jt,Lee:2007wr,Oh:2009zj}.
In extreme environments, such as high temperature and/or density, chiral symmetry breaking should be also modified, and it would strongly affect the chiral partner structures of diquarks and the related baryon spectra.

\section*{Acknowledgments}
We thank Masayasu Harada and Yan-Rui Liu for useful discussions.
This work was supported in part by JSPS KAKENHI Grants No. JP18H05407 (E.H.), No. JP19H05159 (M.O.), and No. JP17K14277 (K.S.).


\bibliography{reference}
\end{document}